\documentclass[conference]{IEEEtran}
\IEEEoverridecommandlockouts
\usepackage{cite}
\usepackage{amsmath,amssymb,amsfonts}
\usepackage{algorithmic}
\usepackage{graphicx}
\usepackage{textcomp}
\usepackage{xcolor}
\usepackage{url}
\usepackage{newunicodechar}
\newunicodechar{≈}{\ensuremath{\approx}}
\usepackage[utf8]{inputenc}
\usepackage{booktabs}
\usepackage{balance}
\usepackage{makecell}
\usepackage{color,soul}
\def\BibTeX{{\rm B\kern-.05em{\sc i\kern-.025em b}\kern-.08em
    T\kern-.1667em\lower.7ex\hbox{E}\kern-.125emX}}
\begin{document}
\DeclareUrlCommand\url{\def\UrlLeft{<}\def\UrlRight{>}\urlstyle{tt}}

\title{Towards Supporting Quality Architecture Evaluation with LLM Tools\\}

\author{\IEEEauthorblockN{Rafael Capilla}
\IEEEauthorblockA{\textit{Rey Juan Carlos University}\\
Madrid, Spain\\
rafael.capilla@urjc.es}
\and
\IEEEauthorblockN{Jorge Andrés Díaz-Pace}
\IEEEauthorblockA{\textit{Universidad del Centro de la Provincia de Buenos Aires}\\
Tandil, Argentina\\
andres.diazpace@isistan.unicen.edu.ar}
\and
\IEEEauthorblockN{Yamid Ramírez}
\IEEEauthorblockA{\textit{Rey Juan Carlos University}\\
Madrid, Spain\\
ye.ramirez.2024@alumnos.urjc.es}
\and
\IEEEauthorblockN{Jennifer Pérez}
\IEEEauthorblockA{\textit{Universidad Politécnica de Madrid}\\
Madrid, Spain \\
jenifer.perez@upm.es}
\and
\IEEEauthorblockN{Vanessa Rodríguez-Horcajo}
\IEEEauthorblockA{\textit{Universidad Politécnica de Madrid}\\
Madrid, Spain \\
vanessa.rodriguez.horcajo@upm.es}
}

\maketitle

\begin{abstract}
Architecture evaluation methods have been extensively used to evaluate software designs. Several evaluation methods have been proposed 
to analyze tradeoffs between different quality attributes. Companies building complex products 
where software is an important part require a qualitative evaluation of the quality aspects demanded by their solutions. Also, having competing qualities leads to conflicts when selecting which quality-attribute scenarios are the most suitable ones for an architecture to tackle. Consequently, the scenarios required by the stakeholders must be prioritized and also analyzed for potential risks. Today, architecture quality evaluation is still carried out manually, often involving long brainstorming sessions to decide on the most adequate quality-attribute scenarios for the architecture. 
To reduce this effort and make the assessment and selection of scenarios more efficient, in this research we propose the use of LLMs to partially automate the evaluation activities. 
As a first step in validating this hypothesis, this paper investigates \texttt{MS Copilot} as an LLM tool to analyze quality-attribute scenarios suggested by students and reviewed by experienced architects. Specifically, our study compares the results of an Architecture Tradeoff Analysis Method (ATAM) exercise conducted in a software architecture course with the results of experienced software architects and with the output produced by the LLM tool. 
Our initial findings reveal that the LLM produces in most cases better and more accurate results regarding risks, sensitivity points and tradeoff analysis of the quality scenarios generated manually, as well as it significantly reduces the effort required for the task. Thus, we argue that the use of generative AI has the potential to partially automate and support architecture evaluation tasks by suggesting more qualitative scenarios to be evaluated and recommending the most suitable ones for a given context.
\end{abstract}

\begin{IEEEkeywords}
Architecture evaluation, quality-attribute scenarios, architecture tradeoffs, Large Language Models
\end{IEEEkeywords}

\section{Introduction}
For more than two decades, software architects have used different architecture evaluation methods (e.g. ALMA, ATAM ARID, SAAM) \cite{clements2001} to assess quality attributes and determine which and how certain quality-attribute properties should be explicitly considered in the architecture before implementing the system. Addressing quality attributes 
requires one to identify risks and non-risks associated to scenarios as well as the impact of sensitivity points on the architecture \cite{cervantes2024}. Furthermore, tradeoffs between competing qualities must be addressed, which often complicate the selection and prioritization of scenarios. This activity often leads to long brainstorming sessions where  architects analyze and select quality scenarios according to the system business goals, among other factors. 

In order to partially automate the evaluation quality scenarios and check if the selected scenarios in an architecture evaluation are adequate, 
we propose the use of generative AI techniques like Large Language Models (LLMs). Our aim is twofold: (i) explore speed-ups for the architecture evaluation process and, (ii) check if LLMs can suggest an accurate evaluation of quality scenarios in terms of risks, sensitivity points and trade-offs. Thereby, making brainstorming sessions more effective, we can assist architects in their evaluation tasks, pinpointing the most suitable scenarios based on their pros and cons, and alerting architects about risks and possible quality-attribute tradeoffs. 

In this work, we investigate how a particular LLM (\texttt{MS Copilot}) can support quality-driven architecture evaluations using a Retrieval Augmented Generation (RAG) strategy \cite{lewis2020retrieval} that feeds the LLM using an architecture report from an undergraduate Computer Science course where students applied the Architecture Tradeoff Analysis Method (ATAM) \cite{clements2001}. Afterwards, two of the authors performed the same exercise to set a ground truth to validate the results which was double-checked by a third co-author. 
Our findings reveal that LLMs often provide more details  than those provided by novice architects (students). However, given the tendency of LLMs to ``always return something'', mechanisms (e.g., a ground truth) to check the LLM outputs with the knowledge of experienced architects become necessary to avoid hallucinations or lack of contextual knowledge. Giving context knowledge and examples to an LLM is important for obtaining more accurate evaluations of the quality-attribute scenarios. Therefore, our results summarize such comparisons and we also provide some lessons lessons for an LLM-assisted evaluation process.

The remainder of this paper is organized as follows. Section II describes the background to understand the relevant concepts, while Section III discusses related works. In Section IV, we outline the study design and research goals, and Section V describes the results of the evaluation performed, followed by a discussion of the main findings in section VI. Finally, Section VII discusses the threats to validity, and  Section VIII presents the conclusions as well as future research paths.

\section{Background}\label{sec:background}
ATAM \cite{ATAM2000} is a manual method that aims to assess the consequences and impact of architectural decisions motivated by quality-attribute requirements. An ATAM team is responsible for analyzing the quality needs of stakeholders and producing the so-called \textit{utility tree}, which is made up of a set of quality scenarios that address specific quality concerns. Such quality concerns are typically described in terms of quality properties or attributes that an architecture and system must satisfy.

Quality attributes can impact different areas of the system and the software architecture as well. In addition, each quality-attribute scenario is characterized by \textit{stimuli} to which an architecture (or system) must respond and a way to measure the quality-attribute response triggered by the stimuli. In this context, the ATAM team must suggest, evaluate, and select the most suitable scenarios by evaluating risks, sensitivity points and tradeoffs between qualities. 

In particular, the tradeoffs for a scenario reflect the tensions between two or more qualities, thus a criterion must be used to prioritize and select among several scenarios during the brainstorming sessions. An experience in analyzing tradeoff points and sensitivity points using the Analytical Hierarchical Process (AHP) method is described in \cite{Ibrahim2009}, where the authors rely on pairwise comparisons to support decision-making of commercial off-the-shelf (COTS) components. The use of AHP to evaluate the quality attributes of the system architecture according to the ISO ISO/IEC 25002:2024 quality model\footnote{\url{https://www.iso.org/standard/78175.html}} is also highlighted in \cite{Darwish2017}. Only a few approaches \cite{Lytra2020} support some kind of automation with respect to architecture quality tradeoffs in multi-attribute decision-making.

Using generative AI and LLMs based on the transformer architecture\cite{Vaswani2017} enable making decisions as a timely alternative to speed up architects' decisions and refine system architectures \cite{Ozkaya2023a}. Not far ago, LLMs were trained and fine-tuned using datasets \cite{brown2020language} to respond to natural language questions. 
Therefore, the use of LLM-based tools with adequate prompts \cite{White2023} is an important challenge, so that architects do not overlook relevant knowledge when evaluating quality scenarios.  Moreover, retrieval augmented generation (RAG) strategies\cite{Singh2025} are commonly used to incorporate a  knowledge context aimed to produce more accurate results when prompting AI assistants. RAG integrates domain-specific sources into an LLM to improve the relevance of the response. Therefore, feeding LLMs with  knowledge about architectural evaluation cases can produce better responses in terms of accuracy when evaluating quality scenarios. 

\section{Related Work}\label{sec:rw}
Nowadays, different works narrate the interplay between software architecture and LLM use \cite{Schmid2025}. 
In \cite{Dhar2024}, the authors investigate the feasibility of LLMs to generate architectural decision records (ADRs) given a specific context. The authors run an exploratory case study to generate ADRs using three different strategies (i.e. zero-shot, few-shot, and fine-tuning) over a number of five repositories containing sample ADRs. 

In \cite{Soliman2025}, the authors evaluate the use of an LLM to answer questions about architectural knowledge (AK) comparing the responses to a predefined ground truth. In their experience, GPT's answers exhibit moderate quality and recall but low precision, in particular finding quality-attribute solutions. This means that the provided answer still requires expert validation. In addition, in a related work \cite{Soliman2025a}, a comparison of seven LLMs supporting architecture knowledge is performed to understand their ability and accuracy to answer AK questions. Other  works focus on reasoning about the decisions made and the paths LLMs can suggest to arrive at the right answer. In \cite{Pace:2024:Design}, the ReAct framework in combination with Monte Carlo Tree Search is used as a reasoning technique through which an architect can trace the reasoning trajectories of the LLM to explore alternative solutions. 

In \cite{Oliveira2025}, the use of LLMs to assess the quality of software architecture diagrams is discussed. The authors evaluated five quality criteria for architecture diagrams in four open-source projects using ChatGPT-4o. However, their results indicate that most of the qualities examined are partially met by the LLM and require expert supervision. 

According to the aforementioned experiences, it becomes clear that, more and more, generative AI techniques and agents can increase the level of automation in software architecture practice \cite{Ivers2025}, assisting researchers and practitioners in manual, time-consuming architecture tasks (e.g. suggest design alternatives with pros and cons, reflect and reasons about design choices, evaluate software architectures). In this content, a typical architecture practice that requires automation is architecture evaluation. 
A recent work discusses the industrial efforts of Volvo using AI in software architecture development \cite{Johansson2025}. The authors present four uses cases at Volvo Contruction Equipment (VCE) in labor-intensive tasks supporting AI-assisted decision-making activities to generate architecture decision records and design solutions. 

\section{Study Design}\label{sec:approach}
The goal of this work is to evaluate whether generative AI can enhance human decision-making and reflective practices \cite{Razavian2020} of novice architects. We analyzed the results of $9$ student projects from a software architecture course in which the participants, organized in teams of $6$ members each, followed an ATAM process to evaluate and suggest a set of quality-attribute scenarios to improve a given architecture. 
We compared the resulting scenarios with the evaluation performed by an LLM. 

Since the outputs of less experienced subjects, like the students, can differ from those of senior architects, two of the co-authors generated and evaluated a reference set of quality-attribute scenarios (ground truth) for the same case study. The selected scenarios were also contrasted with those of the LLM. Thus, we analyzed the students' results with respect to those of an LLM but also with the judgments of senior architects. 

\subsection{Research questions}
To achieve the proposed goal we adopted a mixed research method \footnote{https://www2.sigsoft.org/EmpiricalStandards/docs/standards} combining an exploratory case study with a qualitative analysis. We address the following research questions:

\textbf{RQ1:} \textit{Can LLMs outperform humans in the evaluation of risks in quality scenarios?}

\textbf{Rationale:} In this question we evaluate if LLMs can identify better and more risks during the evaluation of quality scenarios produced by humans. 

\textbf{RQ2:} \textit{Can LLMs outperform humans in the evaluation of sensitivity points in quality scenarios?}

\textbf{Rationale:} In this question we evaluate if LLMs can identify better and more sensitivity points in the architecture during the evaluation of quality scenarios produced by humans. 

\textbf{RQ3:} \textit{Can LLMs outperform humans in architectural trade-offs in quality scenarios?}

\textbf{Rationale:} In this question we evaluate if LLMs can perform the quality attribute trade-off analysis better than humans and also identify possible new trade-offs.

For the three RQs, we seek to understand if these manual evaluation tasks can be improved in terms of accuracy and effort reduction using LLMs.

\subsection{Materials}
We gave the students the same case study consisting of a software architecture represented by UML class-package and deployment diagrams. This architecture represents the solution for migrating from a three-tier monolithic to a microservices system, where the functionality involves six modules (i.e., customers, orders, delivery and routes, payment, statistics and incidents) that have different criticality for the business. The system uses two SQL databases and HTTP/REST as a communication protocol. 

\subsection{Experiment workflow}
In the following, we describe the steps that were carried out with students and researchers. 

\textbf{Activities with students:} The participants performed the ATAM process for four weeks as follows: 

(i) the students used a software architecture designed in a previous assignment as the main input. For each team, one student acting as the customer indicated a set of quality needs to incorporate into the architecture

(ii) the quality needs were discussed and refined with $5$ students acting as the evaluation team

(iii) the students created a utility tree describing quality scenarios for each of the quality attributes derived from the system's quality goals 

(iv) each quality scenario was evaluated to identify possible risks, sensitivity points. In addition, the students had to identify possible tradeoffs between quality attributes affecting the different scenarios

(v) the final scenarios were prioritized according to their importance and finally selected to be depicted in the architecture. 

All the information from the students' assignments was gathered in PDF documents. We fed these documents into \texttt{MS Copilot}\footnote{https://copilot.microsoft.com/} and used a RAG strategy \cite{lewis2020retrieval} to evaluate the degree to which the scenarios chosen by the students were adequate. 

\textbf{Activities with researchers:} As architecture experts, two of the co-authors repeated the same experiment performed by the students and a third co-author reviewed the accuracy of the results. Also, we used \texttt{MS Copilot} to assess the quality of the quality scenarios evaluated by the researchers. 

\textbf{Comparison of results:} We performed a comparison of scenarios evaluated and selected by the students with those produced by the researchers. Also, we compared these results with the assessment provided by the LLM. 

\begin{figure}[ht]
\centering
\includegraphics[width=0.50\textwidth]{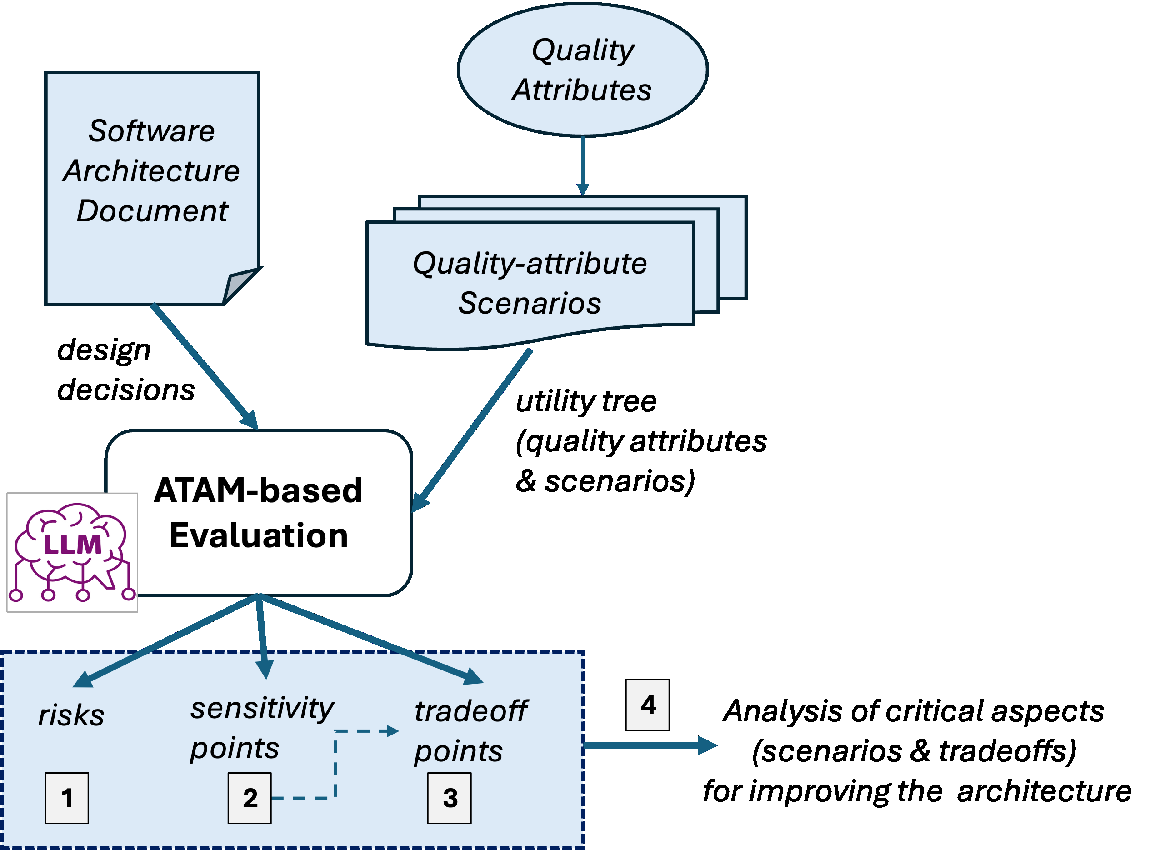}
\caption{Inputs, outputs and steps of the ATAM process. LLM prompts were used at steps $1$, $2$, $3$, $4$ when using \texttt{MS Copilot}.} 
\label{fig:llm-atam} 
\end{figure}

The ATAM process steps to evaluate and select the quality-attribute scenarios are depicted in Figure \ref{fig:llm-atam}. In the ATAM context, we refer to a risk as a problematic architectural decision that can potentially lead to negative consequences if not addressed properly. A sensitivity point, in turn, is a design decision that affects the achievement of a particular quality-attribute scenario. At last, a tradeoff is a design decision that affects several quality-attribute scenarios. 
In addition to having a document with the architectural description, we provided a specialized prompt for \texttt{MS Copilot} at each process step. 
At the end of the process, we asked \texttt{MS Copilot} to perform a selection of scenarios based on the previous information. From the $9$ student groups evaluated, we discarded two groups because the quality of the scenarios was low and they provided several unclear scenarios.  

\subsection{Data Analysis}
As part of our mixed method, we performed a qualitative analysis of the results in terms of completeness and accuracy. We compared the number and quality of the scenarios produced by each team of students with those yielded by the researchers. In this context, we aimed to identify the following aspects: (i) which scenarios are more suitable to meet each quality attribute, (ii) how accurate the scenarios reported by the students are compared to those from the researchers, (iii) how many scenarios per quality attribute the students selected compared to those from the researchers, (iv) how many risks and sensitivity points were identified by researchers and students, and (v) how accurate the trade-offs given by researchers are compared to those reported by the students and the LLM as well.

In addition, we repeated the same analysis comparing the previous results with those reported by the LLM. Therefore, we evaluated which of the solutions is more complete and accurate. Finally, we estimate the effort taken by the students, the researchers, and the use of LLM for the same tasks. All these results are discussed in the next section.
A reproducibility kit with all the artifacts and prompts is provided.

\section{Results}\label{sec:evaluation}
In this section we describe the results of our LLM-approach to do the ATAM process. First, we describe the part performed by the students and then by the researchers. In both cases, we include the LLM evaluation and compare performance. Also, we discarded two groups out of nine because low quality of their results of the utility tree and quality evaluation of the target system. 

\subsection{Evaluation with students}
\textbf{Identification of risks and sensitivity points:} In steps $1$ and $2$, \texttt{MS Copilot} used the prompts to infer risks and sensitivity points related to all the scenarios described by the students. 
The results are shown in Table \ref{tab:risk-sensitivity}. Bold text indicates shared items between the students and the LLM. 
As we can observe, \texttt{MS Copilot} was extremely useful in detecting additional risks and sensitivity points (i.e., items not identified by the students), which are relevant for the selection of scenarios. 
In several groups, there was a good intersection (e.g., $G6$) while in other groups, the LLM detected more risks and sensitivity points (e.g., $G1$, $G3$, $G8$). In those cases where \textit{MS Copilot} suggested additional risks, we list such scenarios in the last column of Table \ref{tab:risk-sensitivity} to indicate that certain scenarios should not have been chosen by the students. 
However, we also acknowledge that this decision depends on factors such as the risk severity, probability of occurrence, or business importance of the scenario, because not all risks are equally critical. If a risk identified for a chosen scenario is not critical, we assume that selecting such scenario would not harm the system. In such situations, \texttt{MS Copilot} could make a risk prioritization that differs from that of humans.

\begin{table*}[ht]
\caption{Identification of risks and sensitivity points (students \& Copilot)}\label{tab:risk-sensitivity}
\scalebox{0.75}{
\begin{tabular}{p{0.10\columnwidth} | p{0.30\columnwidth} | p{0.30\columnwidth}|p{0.70\columnwidth} |p{0.8\columnwidth}| p{0.2\columnwidth}|}
\toprule
Groups & Scenarios with risks (students) & Scenarios with risks (Copilot)  & Sensitivity points (students) & Sensitivity points (Copilot) & Scenarios selected \\
\hline
G1 & \textbf{1.1}, \textbf{1.4}, \textbf{3.1,} \textbf{4.1} & \textbf{1.1}, \textbf{1.4,} 2.2, 2.3, \textbf{3.1}, 3.3, \textbf{4.1}, 4.4, 6.3 & \textbf{Database}, \textbf{backup security}, \textbf{database scalability} & \textbf{Database}, Backup \& restore, Encryption key, Kubernetes failover, Cloud backup, API Gateway, Stripe connect dependency & 3.3  \\
\hline
G2 & \textbf{1.1}, 1.3, \textbf{1.4}, \textbf{2.1}, \textbf{2.4}, \textbf{3.1}, \textbf{3.3}, \textbf{3.4}, 3.5 & \textbf{1.1}, 1.2, \textbf{1.4}, 1.5, \textbf{2.1}, 2.3, \textbf{2.4}, \textbf{3.1}, \textbf{3.3}, \textbf{3.4} & None & Data layer availability, real-time capacity, Backups, Kubernetes autoscaling, Operational monitoring, Microservices decomposition &1.2, 2.3, 3.3\\
\hline
G3 & \textbf{1.1}, \textbf{1.2}, \textbf{1.3}, \textbf{1.4}, \textbf{1.5}, \textbf{2.1}, \textbf{2.2}, \textbf{2.3}, \textbf{3.1}, \textbf{3.2}, \textbf{3.3}, \textbf{3.4}, \textbf{4.1}, \textbf{4.2}, \textbf{4.3} & \textbf{1.1}, \textbf{1.2}, \textbf{1.3}, \textbf{1.4}, \textbf{1.5}, \textbf{2.1}, \textbf{2.2}, \textbf{2.3}, \textbf{3.1}, \textbf{3.2}, \textbf{3.3}, \textbf{3.4}, \textbf{4.1}, \textbf{4.2}, \textbf{4.3} & \textbf{Peak workload}, \textbf{latency}, \textbf{bottlenecks}, \textbf{database security}, \textbf{biometric security} & \textbf{Database availability}, \textbf{Database encryption}, \textbf{Load balancer}, Algorithmic scalability, Web authentication & 1.1, 2.3, 3.2\\
\hline
G4 & \textbf{1.1}, \textbf{1.6}, \textbf{2.1}, \textbf{2.4} & \textbf{1.1}, 1.2, \textbf{1.6}, \textbf{2.1}, \textbf{2.4}, 3.1 & \textbf{Database} & \textbf{Database}, encryption system, load balancer, cache, Authentication 2FA & 1.1, 1.6 \\
\hline
G6 & \textbf{1.1}, \textbf{2.2}, \textbf{2.4}, \textbf{3.2} & \textbf{1.1}, \textbf{2.2}, \textbf{2.4}, \textbf{3.2} & \textbf{Database} & \textbf{Database}, Server logic failure & None \\
\hline
G7 & \textbf{1.1}, \textbf{1.2}, \textbf{3.1}, \textbf{3.2} & \textbf{1.1}, \textbf{1.2}, 2.3, \textbf{3.1}, \textbf{3.2} & \textbf{Database encryption}, \textbf{database scalability} & \textbf{Database encryption}, Database availability, Cloud DB capacity, Gateway bottleneck, Statistic module & 3.2 \\
\hline
G8 & \textbf{1.1}, \textbf{1.2}, \textbf{1.3}, \textbf{2.2}, \textbf{2.3}, \textbf{3.1}, \textbf{3.6} & \textbf{1.1}, \textbf{1.2}, \textbf{1.3}, \textbf{2.2}, \textbf{2.3}, \textbf{3.1}, 3.4, 3.5, \textbf{3.6} & Server availability, \textbf{database caching}, \textbf{database encryption} & \textbf{database encryption} Apache Kafka queue, Redis cache, statistics & 1.3, 2.2, 3.1, 3.4 \\
\hline
\end{tabular}
}
\end{table*}

\textbf{Tradeoff analysis:} In step 3, we prompted \texttt{MS Copilot} to perform a quality-attribute tradeoff analysis for the scenarios identified in the utility tree by each group. The results are shown in Table \ref{tab:tradeoff}. In the second and third columns we indicate the qualities involved in each tradeoff identified by the students and by \texttt{MS Copilot}, while the fourth and fifth columns indicate the scenarios involved and selected during the tradeoff analysis by the students and \texttt{MS Copilot} as well. As can be seen, \texttt{MS Copilot} identified more trade-offs between qualities in all cases. Therefore, the analysis performed by the LLM involves more scenarios for the tradeoff analysis. Also, in general terms, the scenarios chosen by \texttt{MS Copilot} are in line with the selection made by the students, except in some cases, where \texttt{MS Copilot} considered that more scenarios could be selected (e.g. $G6$, $G7$). 

\begin{table*}[ht]
\caption{Analysis of tradeoff points (students  \& Copilot)}\label{tab:tradeoff}
\vspace{0.1cm}
\scalebox{0.75}{
\begin{tabular}{p{0.10\columnwidth} | p{0.75\columnwidth} | p{0.70\columnwidth} |p{0.45\columnwidth} |p{0.45\columnwidth}|}
\toprule
Groups & Tradeoff points (students) & Tradeoff points (Copilot)  & Scenarios affected by tradeoffs & Selected scenarios\\
\hline
G1 & \textbf{Performance-Security}, \textbf{Performance-Reliability} & \textbf{Performance-Security,} \textbf{Performance-Reliability}, Security-Usability, Performance-Scalability, Security-Maintainability & Students: None  Copilot: 1.1, 1.2, 2.2, 2.3, 2.4, 4.1, 4.2, 5.2 & Students:\textbf{1.2}, \textbf{2.1}, \textbf{3.3}, \textbf{4.2}, 5.2, 6.6 Copilot: \textbf{1.2,} \textbf{2.1}, \textbf{3.3}, \textbf{4.2}, 5.2 \\
\hline
G2 & \textbf{Security-Performance}, \textbf{Security-Availability} & \textbf{Security-Performance}, \textbf{Security-Availability}, Performance-Availability & Students: \textbf{1.1}, \textbf{1.3}, 1.6 Copilot: \textbf{1.1}, \textbf{1.3}, 2.1, 2.3, 2.4, 2.6, 3.1, 3.6 & Students: \textbf{1.2}, \textbf{1.6}, \textbf{2.3}, \textbf{2.6}, \textbf{2.9}, 3.3, 3.6 Copilot: \textbf{1.2}, \textbf{1.6}, \textbf{2.3}, \textbf{2.6}, \textbf{2.9}, 3.6 \\
\hline
G3 & Scalability-Performance, \textbf{Security-Performance}, Availability-Performance, Authenticity-Compatibility & Scalability-Latency, \textbf{Security-Performance}, Availability-Complexity, Authenticity-Usability & Students: \textbf{All}  Copilot: \textbf{All} & Students: 1.1, \textbf{2.3}, 3.2, 4.4 Copilot: 1.2, \textbf{2.3}, 3.1, 4.3 \\
\hline
G4 & \textbf{Security-Performance}, \textbf{Scalability-Performance} & \textbf{Security-Performance}, \textbf{Scalability-Performance}, Scalability-Complexity, Performance-Consistency, Security-Usability, Availability-Complexity & Students: 1.1, 1.6, 2.2, 2.3, \textbf{3.1}  Copilot: 1.2, 2.1, 2.4, 2.5, \textbf{3.1} & Students: 1.1, 1.6, 2.2, 2.5, \textbf{3.1} Copilot: 1.2, 2.1, 2.4, \textbf{3.1} \\
\hline
G6 & Performance-Interoperability, \textbf{Security-Performance}, Performance-Availability & S\textbf{ecurity-Performance}, Interoperability-Complexity, Availability-Cost, Scalability-Simplicity & Students: \textbf{All}  Copilot: \textbf{All} & Students: \textbf{1.2}, \textbf{2.1}, 3.4  Copilot: \textbf{1.2}, \textbf{2.1}, 2.3, 3.4 \\
\hline
G7 & \textbf{Security-Performance} & \textbf{Security-Performance}, Performance-Scalability, Scalability-Cost, Security-Usability, Performance-Complexity, Scalability-Maintainability & Students: \textbf{1.1}, \textbf{1.2}, \textbf{1.3}  Copilot: \textbf{All} & Students: \textbf{1.3}  Copilot: 1,2, \textbf{1.3}, 2.3, 3.1, 3.3 \\
\hline
G8 & Reliability-Availability, \textbf{Reliability-Performance-Efficiency}, \textbf{Security-Performance} & \textbf{Reliability-Performance-Efficiency}, \textbf{Security-Performance}-Efficiency, Scalability-Maintainability, Availability-Cost, Security-Usability & Students: \textbf{All}  Copilot: \textbf{1.1}, \textbf{1.2}, \textbf{1.3}, \textbf{2.1}, \textbf{2.2}, \textbf{2.3}, \textbf{3.1}, \textbf{3.2}, \textbf{3.3}, \textbf{3.4} & Students: 1.3, \textbf{2.1}, 2.2, 2.4 Copilot: 1.2, 1.3, \textbf{2.1}, 3.1 \\
\hline
\end{tabular}
}
\end{table*}

\textbf{Selection of scenarios:} For step 4, Table \ref{tab:atam} shows, for each group and quality attribute, the identifiers of the scenarios selected by the students and by \texttt{MS Copilot}. Red text marks those cases in which the scenarios chosen by the students and by the LLM were very different, whereas green text marks those that exhibited a very good match. In the rest of the cases, there was some degree of scenario matching, but also some differences between the students and \texttt{MS Copilot}. For $G6$, we noticed the matching of the scenarios was almost perfect. In general, only in $4$ out of $30$ cases, the scenarios selected by \texttt{MS Copilot} diverged from the students' choices. Thus, overall, we can consider that the use of an LLM was beneficial to evaluate and select scenarios. Finally, we asked \texttt{MS Copilot} to select the most suitable scenarios for a specific quality attribute and area of the system affected by it, but the LLM often selected only one scenario. This is an example of the  challenges that still exist  
when asking an LLM to perform this kind of tasks. 

\begin{table*}[ht]
\caption{Quality-attribute scenarios selected (students)}\label{tab:atam}
\vspace{0.1cm}
\scalebox{0.75}{
\begin{tabular}{p{0.25\columnwidth} | p{0.30\columnwidth} | p{0.30\columnwidth} |p{0.30\columnwidth} |p{0.30\columnwidth} | p{0.30\columnwidth} |p{0.30\columnwidth} |p{0.25\columnwidth} |}
\toprule
Groups and scenarios & Quality 1 & Quality 2  & Quality 3 & Quality 4 & Quality 5 & Quality 6 & Quality 7\\
\hline
G1 & Security & Security & Reliability & Performance Efficiency (data access) & Performance Efficiency (orders) & & \\
Sc. students & \textcolor{green}{1.2} & \textcolor{green}{1.2} & \textcolor{green}{3.3, 4.2} & \textcolor{green}{5.2} & \textcolor{red}{6.2} & &\\
Sc. Copilot & \textcolor{green}{1.2} & \textcolor{green}{2.1} & \textcolor{green}{3.3, 4.2} & \textcolor{green}{5.2} & \textcolor{red}{None} & &\\
\hline
G2 & Security (database) & Security (server) & Performance (database) & Performance (server) & Performance (orders) & Availability (database) & Availability (orders)\\
Sc. students & 1.2 & 1.6 & 2.3 & \textcolor{red}{2.6} & 2.9 & \textcolor{red}{3.3} & 3.6\\
Sc. Copilot & 1.2, 1.3 & 1.5, 1.6 & 2.2, 2.3 & \textcolor{red}{2.5} & 2.8, 2.9 & \textcolor{red}{3.1, 3.2} & 3.4, 3.6\\
\hline
G3 & Scalability & Security (database) & Security-Authenticity & Availability & &  & \\
Sc. students & 1.2 & 2.3 & 3.2 & \textcolor{red}{4.1, 4.2} & & &\\
Sc. Copilot & 1.1, 1.2 & 2.2, 2.3 & 3.1, 3.2 & \textcolor{red}{4.3, 4.4} & & &\\
\hline
G4 & Security (database) & Security (client) & Scalability (GW) & Scalability (database) & Performance (database)  & & \\
Sc. students & \textcolor{green}{1.1j} & \textcolor{red}{1.6} & \textcolor{green}{2.2} & \textcolor{green}{2.5} & \textcolor{green}{3.1} & &\\
Sc. Copilot & \textcolor{green}{1.1} & \textcolor{red}{1.4} & \textcolor{green}{2.2} & \textcolor{green}{2.5} & \textcolor{green}{3.1} & &\\
\hline
G6 & Interoperability & Security & Availability &  &   & &\\
Sc. students & \textcolor{green}{1.2} & \textcolor{green}{2.1} & \textcolor{green}{3.4} & & & &\\
Sc. Copilot & \textcolor{green}{1.2} & \textcolor{green}{2.1} & \textcolor{green}{3.4} & & & &\\
\hline
G7 & Security & Performance & Scalability &  &  & &\\
Sc. students & 1.3 & \textcolor{green}{2.2, 2.3} & 3.1 & & & &\\
Sc. Copilot & 1.2, 1.3 & \textcolor{green}{2.2, 2.3}  & 3.2, 3.3 & & & &\\
\hline
G8 & Reliability & Performance Efficiency & Security &  &  & &\\
Sc. students & \textcolor{green}{1.3} & 2.1, 2.2 & 3.1, 3.2, 3.4 & & & &\\
Sc. Copilot & \textcolor{green}{1.3} & 2.1  & 3.2, 3.3 & & & &\\
\hline
\end{tabular}
}
\vspace{-0.5cm}
\end{table*}

\subsection{Evaluation by architecture experts}
Two co-authors acting as experienced software architects (with more than 20 years doing research and professional activities in the software architecture field) repeated similar tasks as performed by the students to identify and select quality scenarios using the same case study. Since the creation of scenarios by these experts could differ from what the students did, and in order to define a ground truth (GT) for comparison with the LLM results, the scenarios produced by the best two groups(G1 and G3) were taken and refined based on the authors' evaluation expertise. These refined scenarios were reviewed by another co-author with vast architecture experience. 
The scenarios were considered as representative examples to double-check the validity of the responses given by both the students and the LLM. Nonetheless, 
we acknowledge that certain scenarios from other groups could have tackled different angles of the case study. 


Due to space constraints, Table \ref{tab:security-scenarios-authors} only shows the refined scenarios for \textit{security}, but analogous tables were produced for \textit{reliability, performance-efficiency and availability} \footnote{Some of the tables with the corresponding scenarios can be found in the replication package}. The \textit{security} scenarios affected two portions of the architecture: the data access layer and the business logic layer. The pairs in the \textit{priority} column indicate the business importance of the scenario and its technical difficulty. Each scenario is summarized under the \textit{mixed scenario} column, and then split into its requirement and decision parts to clarify the problem and solution (i.e., decision) aspects. Note that the students did not make this distinction in their reports, although we believe this separation of concerns provides a better context for scenario selection.

\begin{table*}[ht]
\caption{Excerpt of Scenarios produced by the expert architects (researchers) in the ground truth}\label{tab:security-scenarios-authors}
\vspace{0.1cm}
\scalebox{0.75}{
\begin{tabular}{p{0.10\columnwidth} | p{0.15\columnwidth} | p{0.10\columnwidth} |p{0.70\columnwidth} |p{0.70\columnwidth}| p{0.65\columnwidth}|}
\toprule
QA & Architecture module & Priority  & Mixed scenario & Requirement part (from scenario) & Decision part (from scenario)\\
\tabularnewline
\hline
Security & Data access layer & (H,H) & R-SC1.1: The AES-256 algorithm can be used to encrypt stored data (i.e., data at rest), ensuring confidentiality and preventing unauthorized access. The key length is 256 bits. The encryption speed is 7,500 bytes/s, and the decryption speed is 10,000 bytes/s. This encryption requires considerable CPU and memory resources & Encrypt stored data (i.e., data at rest) ensuring confidentiality and preventing unauthorized access. & Use AES-256 for data-at-rest encryption. Key length is 256 bits. Encryption speed is 7,500 bytes/s, while decryption speed is 10,000 bytes/s. This algorithm consumes considerable CPU and memory resources.
\\
\hline
&  & (H,M) & \textcolor{green}{R-SC 1.2: To encrypt stored data (i.e., data at rest), the pgcrypto extension with AES-256 can be used, ensuring confidentiality and preventing unauthorized access. The key length is 256 bits. The encryption speed is 7,500 bytes/s, and the decryption speed is 10,000 bytes/s. This type of encryption can be applied transparently, without the need to modify the application logic, with the advantage of performing encryption and decryption directly in the database} & Stored data (i.e., data at rest) must be encrypted to ensure confidentiality and prevent unauthorized access. The encryption mechanism should be transparent to existing application code, ideally requiring no or minimal changes. Encryption/decryption may be performed close to the data store (e.g., in the DB layer) to centralize enforcement & Use PostgreSQL’s pgcrypto extension with AES-256. Key length 256 bits, with measured speeds 7,500 / 10,000 B/s for encryption and decryption respectively.	Encryption/decryption are implemented as functions performed inside the DB engine, thus the functionality is kept separate from the application logic.\\
\hline
& & (M,M) & R-SC 1.3: The AES-128 algorithm can be used to encrypt stored data, ensuring confidentiality and preventing unauthorized access. The key length is 128 bits. The encryption speed is 10,000 bytes/s, and the decryption speed is 15,000 bytes/s & Stored data must be encrypted to ensure confidentiality and prevent unauthorized access. A balance between strong encryption and higher performance/throughput is necessary & Use AES-128 for data-at-rest encryption. Key length is 128 bits. Observed throughput is 10,000 B/s for encryption, and 15,000 B/s decrypt for decryption. This option is faster than the AES-256 option.
\\
\hline
& & (M,L) & R-SC 1.4: The ChaCha20 algorithm can be used to encrypt stored data, ensuring confidentiality and preventing unauthorized access. The key length is 256 bits. The encryption and decryption speed is 60,000 bytes/s. This type of encryption requires 20 rounds of nonlinear functions to encrypt information in a more complex manner & Stored data (i.e., data at rest) must be encrypted to ensure confidentiality and prevent unauthorized access. There is a need for very high throughput in encryption/decryption. The chosen method must provide strong modern security guarantees & Use the ChaCha20 stream cipher algorithm, with 256-bit key and 20 rounds. Achievable speed is 60,000 B/s for both encryption and decryption, which is significantly higher than the previous AES numbers.
\\
\hline
& Business logic layer & (H,H) & \textcolor{green}{R-SC 2.1: Authentication can be guaranteed using JWT (JSON Web Tokens). This allows user identity to be validated and sessions to be controlled without storing server state. Token generation takes approximately 0.5 ms per token, with validation time of 0.3 ms. If necessary, token expiration can be configured} & The system must authenticate users and control sessions. The session mechanism should be stateless on the server side (no per-session server-side storage). Token issuance and validation must be fast enough to avoid degrading overall response time. Token expiration must be configurable to balance security and usability & Use JWT (JSON Web Tokens) for representing authenticated identity and session state. Measured performance is ~0.5 ms to generate a token, and ~0.3 ms to validate it. Use JWT expiry fields to control token lifetime.
\\
\hline
& & (H,M) & R-SC 2.2: Authentication can be guaranteed using OAuth 2.0 by implementing access tokens with defined expiration times (15 minutes) and refresh tokens to extend the session. The average latency of the authentication process is 250 ms &  The system must provide user authentication (and typically delegated authorization) using a token-based mechanism. Access tokens should be short-lived (e.g., 15 minutes) for security. Refresh tokens should allow extending the session without requiring full re-authentication each time. The end-to-end authentication flow should have bounded latency (the solution currently achieves 250 ms on average) & Use OAuth 2.0 flows (e.g., Authorization Code, etc.). Configure access tokens with 15-minute expiration.Use refresh tokens for session continuation. Actual implementation achieves ~250 ms average authentication latency.
\\
\hline
& & (M,H) & R-SC 2.3: Authentication can be guaranteed through facial recognition biometrics using libraries such as OpenCV. This allows users' identities to be verified by capturing and validating their facial image. The average latency of the authentication process is 200 ms & The system must authenticate users using a biometric factor (specifically, face recognition). Authentication must capture and validate a live facial image. End-to-end biometric authentication should meet a given latency requirement so that login does not disrupt user experience & Use facial recognition as the authentication method. Use OpenCV (or a similar library) to process and validate facial images. Achieved latency is ≈ 200 ms per authentication.
\\
\hline
& & (H,M) & R-SC 2.4: Authentication can be guaranteed through mutual authentication between clients and servers using X.509 certificates. This method guarantees identity using asymmetric keys and a certificate signed by a Certification Authority (CA). The average latency of the authentication process is 150 ms & The system must authenticate both clients and servers (i.e., mutual authentication) to prevent impersonation attacks. Authentication must be based on asymmetric cryptography for strong identity assurance. The authentication handshake must be fast enough to avoid degrading connection establishment & Use X.509 certificates for identity. Certificates must be CA-signed. Mutual authentication is implemented via a protocol like TLS with client certificates. Observed latency is ≈ 150 ms.
\\
\hline
\end{tabular}
}
\end{table*}

\textbf{Identification of risks and  sensitivity points:} For each scenario, we identified the risks and sensitivity points taking into account the architectures used by the students. The corresponding risks are shown in Table \ref{tab:R-SP-Tradeoff-authors}. When it comes to the sensitivity points, 
we identified the following ones: \textit{security breach and/or- failure of the database, security breach and/or failure of the payment system, failure of the strategy algorithm selecting the delivery routes.} As can be seen, these are fewer risks and sensitivity points than those reported by the students and later enriched by the LLM.

\textbf{Tradeoff analysis:} Based on the scenario descriptions that we used as ground truth, we performed a tradeoff analysis to identify competing forces among them. The results of this analysis are shown in Table \ref{tab:R-SP-Tradeoff-authors}. 

\textbf{Selection of scenarios:} Finally, according to the previous risks, sensitivity and tradeoff points, we selected the scenarios we felt most adequate for the quality requirements. The selected scenarios are marked in green in Table \ref{tab:security-scenarios-authors}. All the information about risks, sensitivity and tradeoff points is summarized in Table \ref{tab:R-SP-Tradeoff-authors}. The aforementioned results are organized around the four quality attributes extracted from the student groups. 


\begin{table*}[ht]
\caption{An excerpt of risks, sensitivity points and tradeoffs selected by researchers}\label{tab:R-SP-Tradeoff-authors}
\scalebox{0.75}{
\begin{tabular}{p{0.40\columnwidth} | p{0.70\columnwidth} | p{0.70\columnwidth} |p{0.70\columnwidth}|}
\toprule
QA & Risks & Sensitivity points & Tradeoff analysis\\
\hline
Security & Weak encription using 128 bits (R-SC 1.3) & Security breach of the database, security breach of the payment system & Security-Performance (R-SC1.1, R-SC.12, R-SC1.3, R-SC1.4, R-SC2.1,  R-SC2.2,  R-SC2.3,  R-SC2.4)\\
\hline
Reliability & No potential risk & Failure of the database, failure of the strategy algorithm selecting the delivery routes & Reliability-Performance (R-SC3.1, R-SC3.2, R-SC3.3); Reliability-Scalability (R-SC3.2); Reliability-Capacity (R-SC3.1, R-SC3.2, R-SC3.3) \\
\hline
Performance-Efficiency & Possible low latency (R-SC4.1) & N/A & Performance-Efficiency-Availability (R-SC4.1,R-SC4.4); Performance-Efficiency-Scalability (R-SC4.1,R-SC4.3)  \\
\hline
Availability & No potential risk & Failure of the database, failure of the strategy algorithm selecting the delivery routes & Availability-Recoverability (R-SC6.1); Availability-Redundancy (R-SC6.1, R-SC6.2) \\
\hline
\end{tabular}
}
\end{table*}

\subsection{Comparison of results}

We performed qualitative and quantitative analyses of the outcomes of both the students and LLM against the ground truth. An example of the outputs generated by \texttt{MS Copilot} using the sequence of prompts from Figure \ref{fig:llm-atam} is given here\footnote{https://anonymous.4open.science/r/archevaluation-llms-B70E/prompting-example.md}.

\subsubsection{Quantitative Analysis:} In this section we assess which risks, sensitivity points and tradeoffs were considered by the students and researchers (in the ground truth), and also which scenarios were finally selected. 

\textbf{Identification of risks and sensitivity points:} The risks identified by the researchers for the scenarios belonging to the four selected qualities (see Table \ref{tab:security-scenarios-authors} and the remaining tables in the replication package) are summarized in Table \ref{tab:R-SP-Tradeoff-authors}. As experts, we did not find many risks in the scenario descriptions of the ground truth. The contrast with the risks stated by the students and also by \texttt{MS Copilot} (Table \ref{tab:risk-sensitivity}) is evident. Several groups pointed out that many scenarios could trigger risks, but they were often not grounded in architectural or contextual information from the system.

Regarding the risks suggested by \textit{MS Copilot}, it amplified the students' analysis, uncovering more risks and adding more details to them. Although there is value in these details, we observed that the LLM tended to suggest risks for almost every scenario without taking into account if those risks were critical or just something negative that could affect a scenario. For this reason, we disregarded several of the LLM-inferred risks. 
Probably, adopting a few-shot strategy and giving more concrete information about the system context  could have improved the accuracy of \texttt{MS Copilot} towards less but critical risks for the analyzed scenarios. 

Many of the sensitivity points reported by the students (Table \ref{tab:risk-sensitivity}) mapped to those of the ground truth, but their descriptions lacked precision. 
For instance, the students only mentioned \textit{database} as a sensitivity point, while in the ground truth we characterized the issue related to the sensitivity point (e.g. a security breach in the database). The description of a sensitivity point is not only a matter of naming the architectural element but also to indicate how it affects the quality of interest. In this regard, the responses from \texttt{MS Copilot} were more precise but still having room for improvement. Prompting the LLM to be precise and explain the rationale behind each sensitivity point could have provided a better answer here. For instance, in some cases, the LLM was helpful by providing a  number of alternatives for a particular question. Despite its assistance, \texttt{MS Copilot} cannot yet substitute for the long experience of architecture experts when it comes to short but precise judgments.

\textbf{Tradeoff analysis:} Like in the previous step, we performed a tradeoff analysis to identify competing qualities in the scenarios used as ground truth. According to Table \ref{tab:R-SP-Tradeoff-authors}, we indicate in the last column a sample of the tradeoffs for the scenarios belonging to the four qualities under analysis. Although the table does not include the tradeoffs for the rest of the scenarios and detailed tradeoff explanation, we believe it provides a representative example that is very similar of what the students did, including the output of the LLM. We identified a number of tradeoffs that shared similarities with those generated by \texttt{MS Copilot}. Thus, we can say that the LLM was able to infer tradeoffs similar to those of experts, which can be helpful for less experienced users like the students. In general, we did not observe significant differences or particular situations where the LLM outperformed the assessment of the researchers.

\textbf{Selection of scenarios:} Most of the qualities selected by the groups mapped to the qualities selected by the researchers, including the architectural portions these qualities affect, as reflected in Table \ref{tab:qualities-r}. However, some qualities are not shown in the table (e.g. security affecting the orders module in group G1, security affecting the database in G3) while others (e.g., scalability affecting the server and identified by group G3) were not considered by the researchers. These disagreements in the qualities are marked in red in the table. 

It might be possible that some qualities indicated by the students and marked in red in the \textit{other qualities} column of Table \ref{tab:qualities-r} could be included in the ground truth. In order to establish a minimal agreement for setting a basis of scenarios, we decided to have a small number of qualities affecting some parts of the system rather than a long list. The qualities chosen by the researchers as ground truth constitute the minimal agreement on the relevant scenarios for the system.
We felt that other qualities (as well as the corresponding scenarios) selected by the groups did not make much sense (e.g. G6: interoperability of the statistics module) and did not include them. Except for these cases, Table \ref{tab:qualities-r} shows that all the qualities selected by the researchers, affecting different portions of the architecture, were indeed considered by several groups.

In addition, Table \ref{tab:security-scenarios-authors} 
shows an excerpt of the scenarios proposed and selected by the researchers for security. The selected scenarios are marked in green, which derived from our assessment of risks, sensitivity points, and tradeoffs. In this particular case, we agreed with the results of group G1 in the selection of two scenarios, but in other cases the choices could differ.
Despite group G1 arrived to the same scenario selection as the researchers for \textit{security}, our results regarding risks, sensitivity points and tradeoffs were more accurate. Compared to group G3 and the same quality, the selection of the first scenarios used the same encryption technique (i.e. AES-256), while other groups like G2 made a decision for AES-192 bits encryption, which is a less stronger algorithm. In general, this kind of comparisons and design decisions is very dependent of the customer's needs. For instance, during the tradeoff analysis, the students compared the performance of this scenario versus AES-256 encryption, and then preferred an algorithm being a bit faster but less secure. 

\begin{table*}[ht]
\caption{Qualities selected by researchers and mapped to qualities selected by students}\label{tab:qualities-r}
\vspace{0.1cm}
\scalebox{0.75}{
\begin{tabular}{p{0.25\columnwidth} | p{0.30\columnwidth} | p{0.30\columnwidth} |p{0.30\columnwidth} |p{0.30\columnwidth} | p{0.30\columnwidth} |p{0.60\columnwidth} |}
\toprule
Researchers & Security (Data access layer) & Security (Business logic layer) & Reliability (Data access layer) & Performance Efficiency (Data access layer, order module) & Availability (Database) & Other qualities\\
\hline
Groups & G1, G4, G6, G7, G8 & G2 & G1 & G1, G2, G4 & G2, G3 & \textcolor{red}{G3 (scalability of the server), G2 (security of the database), G2, G8 (performance of the server), G2, G6 (availability of the server), G4 (security of the client), G4 (scalability of the API Gateway), G4, G7 (scalability of the database), G6 (interoperability of the statistics module), G7 (performance of the delivery algorithm)} \\
\hline
\end{tabular}
}
\vspace{-0.5cm}
\end{table*}

\subsubsection{Quantitative Analysis}
Much of the comparisons so far involving students, researchers and the use of \texttt{MS Copilot} rely on a qualitative assessment of the results, which introduces a degree of subjectivity. To reduce possible biases and achieve more systematization in the comparisons, we used a Likert scale to categorize both the severity of risks and the importance of sensitivity points. The scale has the following values pairs:
\texttt{0: not important at all / no risk, 1: somehow important / some risk, 2: very important / important risk, 3: essential / high risk}. The first value of the pair refers to a sensitivity point and the second one to a risk. 

For the risks, instead of listing the scenarios in which the students and the LLM identified risks, we summarized them as a short list.  As a result, we describe the ratings in Table \ref{tab:severity-R-SP}, which are marked in blue using square brackets. We categorized the risks and sensitivity points reported by the students, while the categorization of those found by the LLM was done by prompting \texttt{MS Copilot}. Although the table does not describe the complete list of risks and sensitivity points with the ratings provided by the LLM, it gives a good summary to compare what the students did and how the LLM behaved. In general terms, LLMs tend to be conservative, identifying more risks and sensitivity points than humans, thus using a scale becomes important to understand its relevancy for a given system. 

\begin{table*}[ht]
\caption{Categorization of the severity of risks and importance of sensitivity points using a likert scale}\label{tab:severity-R-SP}
\scalebox{0.75}{
\begin{tabular}{p{0.10\columnwidth} | p{0.60\columnwidth} | p{0.55\columnwidth} |p{0.50\columnwidth} |p{0.70\columnwidth}|}
\toprule
Groups & Scenarios with risks (students) & Scenarios with risks (Copilot)  & Sensitivity points (students) & Sensitivity points (Copilot) \\
\hline
G1 & Encryption performance\textcolor{blue}{[1]}, Server / cloud bottlenecks\textcolor{blue}{[2]} Restore time\textcolor{blue}{[0]} & Encryption performance\textcolor{blue}{[3]}, Blocking backups\textcolor{blue}{[3]}, Business logic authentication\textcolor{blue}{[2]}, Throughput bottlenecks\textcolor{blue}{[2]}. Adequate storage\textcolor{blue}{[1]}, Restore downtime\textcolor{blue}{[1]}, Data access caching latency\textcolor{blue}{[3]} & Database\textcolor{blue}{[3]}, backup security\textcolor{blue}{[2]}, database scalability\textcolor{blue}{[3]} & Database\textcolor{blue}{[3]}, Backup \& restore\textcolor{blue}{[3]}, Encryption key\textcolor{blue}{[3]}, Kubernetes failover\textcolor{blue}{[2]}, Cloud backup\textcolor{blue}{[2]}, API Gateway\textcolor{blue}{[1]}, Stripe connect dependency\textcolor{blue}{[1]}\\
\hline
G2 & Encryption performance\textcolor{blue}{[1]}, Database replication\textcolor{blue}{[1]}, Concurrency of users\textcolor{blue}{[1]} & Weak cryptography\textcolor{blue}{[2]}, Encryption performance\textcolor{blue}{[2]}, Poor scalability\textcolor{blue}{[3]}, Failover orchestration\textcolor{blue}{[1]}, Data consistency\textcolor{blue}{[2]}, Process communication\textcolor{blue}{[0]} & None & Data layer availability\textcolor{blue}{[3]}, real-time capacity\textcolor{blue}{[3]}, Backups\textcolor{blue}{[2]}, Kubernetes autoscaling\textcolor{blue}{[2]}, Operational monitoring\textcolor{blue}{[2]}, Microservices decomposition\textcolor{blue}{[1]} \\
\hline
G3 & Network traffic\textcolor{blue}{[1]}, Configuration optimization\textcolor{blue}{[0]}, Requests latency\textcolor{blue}{[1]}, Server performance\textcolor{blue}{[2]}& Server bottlenecks\textcolor{blue}{[2]}, Database encryption\textcolor{blue}{[2]}, Login failure\textcolor{blue}{[1]}, Database inactivity\textcolor{blue}{[2]} & Peak workload\textcolor{blue}{[2]}, latency\textcolor{blue}{[2]}, bottlenecks\textcolor{blue}{[3]}, database security\textcolor{blue}{[3]}, biometric security\textcolor{blue}{[2]} & Database availability\textcolor{blue}{[3]}, Database encryption\textcolor{blue}{[3]}, Load balancer\textcolor{blue}{[2]}, Algorithmic scalability\textcolor{blue}{[1}], Web authentication\textcolor{blue}{[1]} \\
\hline
G4 & Encryption performance\textcolor{blue}{[1]}, Latency using large security tokens\textcolor{blue}{[1]} & Balancer scalability\textcolor{blue}{[2]}, Database vulnerability\textcolor{blue}{[2]}, Cyphering delay\textcolor{blue}{[1]}, Biometric authentication\textcolor{blue}{[2]}, Database inactivity\textcolor{blue}{[3]} & Database\textcolor{blue}{[3]} & Database\textcolor{blue}{[3]}, encryption system\textcolor{blue}{[2]}, load balancer\textcolor{blue}{[2]]}, cache\textcolor{blue}{[1]}, Authentication 2FA\textcolor{blue}{[1]} \\
\hline
G6 & Concurrency of users\textcolor{blue}{[1]}, Cryptography robustness\textcolor{blue}{[3]} & Performance decay\textcolor{blue}{[1]}, Cryptography robustness\textcolor{blue}{[3]}, Database compatibility \textcolor{blue}{[2]}, Replication inconsistencies\textcolor{blue}{[2]} & Database\textcolor{blue}{[3]} & Database\textcolor{blue}{[3]}, Server logic failure\textcolor{blue}{[3]} \\
\hline
G7 & Cryptography robustness\textcolor{blue}{[3]}, Database scalability\textcolor{blue}{[2]}, Vendor lock-in\textcolor{blue}{[2]} & Cryptography robustness\textcolor{blue}{[2]}, Database scalability\textcolor{blue}{[2]}, Vendor lock-in\textcolor{blue}{[1]}, Performance pf the delivery algorithm\textcolor{blue}{[1]} & Database encryption\textcolor{blue}{[3]}, database scalability\textcolor{blue}{[2]} & Database encryption\textcolor{blue}{[3]}, Database availability\textcolor{blue}{[3]}, Cloud DB capacity\textcolor{blue}{[2]}, Gateway bottleneck\textcolor{blue}{[1]}, Statistic module\textcolor{blue}{[0]} \\
\hline
G8 & Server monitoring\textcolor{blue}{[0]}, Data synchronization\textcolor{blue}{[1]}, Caching failure\textcolor{blue}{[3]}, Caching performance\textcolor{blue}{[2]}, Encryption performance\textcolor{blue}{[2]} & Server automatic restart\textcolor{blue}{[2]}, Mirror server synchronization\textcolor{blue}{[1]}, Messaging queue bottlenecks\textcolor{blue}{[2]}, Cache latency\textcolor{blue}{[2]} & Server availability\textcolor{blue}{[3]}, database caching\textcolor{blue}{[1]}, database encryption\textcolor{blue}{[3]} & database encryption\textcolor{blue}{[3]}, Apache Kafka queue\textcolor{blue}{[2]}, Redis cache\textcolor{blue}{[1]}, statistics\textcolor{blue}{[0]} \\
\hline
\end{tabular}
}
\end{table*}

\section{Discussion }\label{sec:discussion}
This section discusses the findings of the study to answer our research questions.

\textbf{RQ1:} \textit{Can LLMs outperform humans in the evaluation of risks in quality scenarios?}

For the risks and sensitivity points identified by the students and \texttt{MS Copilot} in Table \ref{tab:risk-sensitivity}, \texttt{MS Copilot} was able to identify additional risks and sensitivity points in most cases affecting more scenarios than those of the students. However, not all those new risks had the same degree of severity. In our analysis, some risks were important, while others can be de-prioritized in the selection of a scenario. The last column of the table indicates scenarios selected by the students but affected by risks identified by the \texttt{MS Copilot}. Furthermore, the risks identified by the students and the LLM were categorized using a Likert scale, as shown in Table \ref{tab:severity-R-SP}. With this information, we answer \textbf{RQ1} arguing that LLMs outperform humans in the generation of candidate risks, but not necessarily in the evaluation activity.

\textbf{RQ2:} \textit{Can LLMs outperform humans in the evaluation of sensitivity points in quality scenarios?}

\texttt{MS Copilot} identified new sensitivity points in the architecture that were apparently missed by the students. This aspect is important in evaluations, as sensitivity points provide ``leads'' for evaluators to identify tradeoffs or hidden risks later on in the process. In addition, the LLM outputs can serve to the students' analysis by including critical system design areas. As we did for risks, we rated the importance of the sensitivity points with a Likert scale. In this case, we answer to \textbf{RQ2} by saying that LLMs seem very helpful regarding the evaluation of sensitivity points (when provided with enough context), as the LLMs augment the evaluator capabilities by identifying more critical points in the architecture.

\textbf{RQ3:} \textit{Can LLMs outperform humans in architectural trade-offs in quality scenarios?}

From the tradeoff analysis in Table \ref{tab:tradeoff}, \texttt{MS Copilot} identified in all cases more tradeoffs than the students. Nevertheless, some tradeoffs belonged to certain qualities (e.g., cost, usability) that the students were told to ignore due to the lack of cost numbers or their evaluation difficulty (e.g., usability). Although the LLM was able to suggest more scenarios affected by these tradeoffs, a deeper analysis would be recommended to assess the accuracy of the tradeoffs. Finally, from the scenarios selected from the tradeoff analysis, our initial assessment reveals that \texttt{MS Copilot} matched the students' results in several cases but suggested alternative scenarios in other cases. For example, in group $G6$, the students reported only one scenario ($1.3$) for a tradeoff between two qualities while \texttt{MS Copilot} indicated $5$ scenarios from tradeoffs between $6$ pairs of qualities. 
In other cases, such as group $G3$, all scenarios except one are different in the tradeoff analysis between the students and \texttt{MS Copilot}, probably because the LLM identified $6$ tradeoffs versus the two tradeoffs reported by the students. The rating of the tradeoffs is more complex as it involves more reasoning and information to decide which competing qualities should be preferred according to the description of each scenario. Furthermore, the decision part is often driven by customer's preferences (e.g. security over performance), which cannot be assessed without additional information. Therefore, in this case we did not use the Likert scale but rather the ground truth that we, as experts, believe it constitutes the baseline to judge the quality of the trade-off analysis performed by the students and reported by the LLM.


As lessons learned, we can summarize the following ones: 

\begin{enumerate}
    \item 
    An LLM provides more detailed information and explanations than the students, and in some case than the experts, identifying also more risks, sensitivity points and tradeoffs.
    \item 
    Using qualitative comparison (like this study) is important to compare (evaluation) results of less experienced and more experience practitioners, so as to validate not only what humans can do but also the LLM outputs and spot possible ``hallucinations''. 
    \item 
    A quantitative scale simplifies the comparison of the items, when a clear criterion to assign the values of the scale is provided. In our experiment, the weights assigned to risks and sensitivity points were reviewed by a third researcher to reduce possible bias. Using a scale serves to discriminate which risks and sensitivity point really matter, as LLMs tend to provide abundant information but they cannot judge (yet) evaluation items like expert architects do. 
    \item 
    LLM hallucinations can be minimized using a few-shot strategy in the prompts and context knowledge, thus reducing the chance of inaccurate responses. Alternatively, an \textit{answer relevancy score metric} \cite{es-etal-2024-ragas} can be used to compare the quality of the LLM responses with respect to the provided knowledge and examples. 
\end{enumerate}

\section{Threats to Validity}\label{sec:threats}
The threats to validity of our study are discussed below.

\textbf{Internal validity:} It refers to the degree of confidence that the causal relationship being tested is not influenced by other factors or variables. In our experience, the natural variability in LLM responses or any "hallucinations" could have influenced the comparison with students and researchers. To mitigate this threat, we tried to feed the LLM with examples and context knowledge (when possible) for a fair comparison of results. Also, the Likert scale helped us to categorize the severity of risks and the importance of sensitivity points when performing the comparisons. 

\textbf{External validity:} It refers to the extent to which results from a study can be generalized. It is challenging to generalize the results of experiments involving an LLM (\texttt{MS Copilot} or any other), as well as results performed by students (as novice architects). The introduction of a ground truth for the comparisons was an attempt to mitigate this threat. 
Nonetheless, as architecture experts, we observed that  an LLM can often provide information and explanations similar to that of experienced architects, but at a faster pace. Although our approach was not tested in an industrial context, we believe the use of LLMs can accelerate the qualitative evaluation of software architectures. 

\textbf{Construct validity:} It checks how well a test measures the concept it was designed to evaluate. In our exploratory study, the students and researchers performed the same tasks using the same materials, and all these artifacts were given to the LLM (along with appropriate prompts) to provide similar results. Certainly, different or better prompts could have led to different results. We suggest that the relevancy score metric can be used to mitigate this threat. We did not perform a postmortem questionnaire with the students to know their perception and benefits of using an LLM to evaluate the scenarios. We plan to perform this kind of user study in the near future.

\textbf{Conclusion validity:} It is a factor that can lead to an incorrect conclusion about our observations. As mentioned above, using an LLM other than \texttt{MS Copilot} or prompting it differently can generate deviations in the results. To mitigate this threat, we trusted the ground truth and assessment of the researchers (as experienced architects) and had the results double-checked by a third researcher. Nonetheless, complementary mechanisms such as a statistical analysis of the results could have been employed.

\section{Conclusion and Future Work}\label{sec:conclusions}\vspace{-0.1cm}
In this paper, we have investigated how LLMs can provide support and improve the outcomes of an ATAM architecture evaluation process. The evaluation was carried out by students of a software architecture course, who played the role of novice architects, and assisted by \texttt{MS Copilot}. Overall, \texttt{MS Copilot} was able to produce detailed system-related information for key phases of ATAM. More specifically, the LLM generated reasonable results for the identification of risks, sensitivity points and tradeoffs, which contributed to improve the outputs of the students' evaluation process. 

These initial results so far confirm our hypothesis about the usefulness of LLMs in the assessment of quality-attribute scenarios, helping humans to uncover additional information to support their architectural reasoning about quality requirements. On the downside, there is still a challenge on how to provide an LLM with the right examples and contextual architectural knowledge.

As future work, we plan to explore the following paths: (i) use LLMs to create and rank quality scenarios, (ii) perform a deeper analysis of the role of concrete risks, sensitivity points and tradeoffs, (iii) use a score metric to evaluate the quality of the (architectural) outputs generated by am LLM, (iv) investigate reflective approaches (e.g., as part of a ReAct cycle) or LLMs with reasoning capabilities (e.g. GPT-5.1), and (v) adapt techniques from agentic AI (e.g., deep research or Large Action Models) to perform architectural tasks related to quality evaluation 
\cite{Shan2025}. 

\bigskip
\noindent\textit{Data availability:}
We provide an anonymized site with the data used in the paper: https://anonymous.4open.science/r/archevaluation-llms-B70E/


\balance

\bibliographystyle{IEEEtran}
\bibliography{references}
\vspace{12pt}
\end{document}